\pdfoutput=1
\documentclass[%
 reprint,
superscriptaddress,
 amsmath,amssymb,
 aps,
 prl, 
floatfix,
]{revtex4-2}

\usepackage{graphicx}
\usepackage{dcolumn}
\usepackage{bm}

\begin{document}

\title{Flow Reversal in Low-Prandtl-Number Convection via Lateral Confinement}

\author{Zhi-Han Wu}
\affiliation{School of Engineering Science, University of Chinese Academy of Sciences, Beijing 101408, China}

\author{Long Chen}
\email{chenlong@ucas.ac.cn}
\affiliation{School of Engineering Science, University of Chinese Academy of Sciences, Beijing 101408, China}

\author{Yan-Wu Cao}
\affiliation{School of Engineering Science, University of Chinese Academy of Sciences, Beijing 101408, China}

\author{Liang Xue}
\affiliation{School of Engineering Science, University of Chinese Academy of Sciences, Beijing 101408, China}

\author{Ming-Zhu Ai}
\affiliation{State Key Laboratory for Strength and Vibration of Mechanical Structures and School of Aerospace Engineering, Xi’an Jiaotong University, Xi’an 710049, China}

\author{Juan-Cheng Yang}
\affiliation{State Key Laboratory for Strength and Vibration of Mechanical Structures and School of Aerospace Engineering, Xi’an Jiaotong University, Xi’an 710049, China}

\author{Ming-Jiu Ni}
\email{mjni@ucas.ac.cn}
\affiliation{School of Engineering Science, University of Chinese Academy of Sciences, Beijing 101408, China}
\affiliation{State Key Laboratory for Strength and Vibration of Mechanical Structures and School of Aerospace Engineering, Xi’an Jiaotong University, Xi’an 710049, China}

\date{\today}

\begin{abstract}
A prevailing consensus holds that flow reversals of the large-scale circulation (LSC) are suppressed in low-Prandtl-number (Pr) fluids, as high thermal diffusivity rapidly dissipates the energy required to fuel the corner-vortex mechanisms. Here, we report Direct Numerical Simulations of liquid metal convection (Pr$=0.029$) revealing that strong lateral confinement defies this consensus, enabling sustained LSC reversals. We show that confinement triggers a ``plume condensation" transition, reorganizing chaotic thermal plumes into highly coherent, quasi-linear structures. A thermal dissipation analysis demonstrates that this coherence drastically reduces heat loss during transport, allowing plumes to deliver sufficient buoyancy to corner vortices to drive reversals. We map a distinct ``island of reversal" in the parameter space, establishing lateral confinement as a control parameter capable of overcoming the stabilizing effects of high thermal diffusivity.
\end{abstract}

\maketitle

The flow reversal of large-scale circulation (LSC) in Rayleigh-Bénard (RB) convection is a fascinating phenomenon in thermal turbulent systems, with broad implications for atmospheric, oceanic, and geomagnetic dynamics \citep{ahlersHeatTransferLarge2009,glatzmaier1999role, gallet2012reversals, araujo2005wind}. The three-dimensional dynamics of such systems are typically extremely complex, with LSC often accompanied by azimuthal drift, twisting modes, and sloshing behavior. By constraining the system to a quasi-2D configuration, it is possible to eliminate these three-dimensional complexities and thereby reveal the core mechanisms of the reversal process \citep{sugiyamaFlowReversalsThermally2010a, chenReducedFlowReversals2020, niReversalsLargescaleCirculation2015a}. In two-dimensional or quasi-2D models, the growth and merger of corner vortices are widely regarded as the key mechanism triggering flow reversals. However, existing studies generally indicate that in low-Prandtl-number (Pr) fluids (such as liquid metals), the reversal of LSC is significantly suppressed \citep{sugiyamaFlowReversalsThermally2010a, yangComplexnetworkModelingReversal2025}. This is mainly attributed to the high thermal diffusivity of the fluid, which rapidly dissipates the heat accumulated in the corner vortices, thereby hindering their growth. This perspective is supported by various theoretical models, including two-dimensional plume dynamics models \citep{araujo2005wind}, stochastic models \citep{niReversalsLargescaleCirculation2015a}, and data-driven predictive models \citep{yangComplexnetworkModelingReversal2025}, none of which have successfully reproduced LSC reversals under low-Pr conditions, consistent with current simulation and experimental results \citep{sugiyamaFlowReversalsThermally2010a}. Quasi-2D flows under strong lateral confinement, however, diverge fundamentally from their unconfined 2D counterparts. Lateral confinement has been shown to profoundly alter thermal plume morphology and global heat transport \citep{huangConfinementInducedHeatTransportEnhancement2013, chongCondensationCoherentStructures2015, huangEffectsGeometricConfinement2016a, chongEffectPrandtlNumber2018}. This naturally leads to a key question: can the regulatory mechanism provided by sidewall confinement, through effective manipulation of thermal plumes, ultimately achieve the reversal of LSC in low-Pr fluids?

In this Letter, we report the first observation of sustained LSC reversals in low-Pr convection (Pr=0.029) enabled by strong lateral confinement. We demonstrate that such confinement induces a ``plume condensation" transition \citep{chongCondensationCoherentStructures2015}, which transforms chaotic thermal plumes into coherent structures capable of bypassing the high-diffusivity barrier. By systematically exploring the parameter space, we delineate a bounded reversal regime, defining the specific conditions under which sustained reversals emerge.

We performed Direct Numerical Simulations (DNS) of the Boussinesq equations in a rectangular cavity with height $H$, length $L$ ($L/H=1$), and width $W$. The study focuses on a low-Pr fluid (Pr$=\nu/\kappa=0.029$) confined within a quasi-2D domain,  where $\nu$ and $\kappa$ are the kinematic viscosity and thermal diffusivity, respectively. To characterize the reversal regime, we systematically varied the confinement aspect ratio $\Gamma=W/H$ ($0.01 \le \Gamma \le 0.1$) and the Rayleigh number Ra$=g\beta\Delta \theta H^3/(\nu\kappa)$ ($3\times10^6 \le$ Ra $\le 1\times10^9$), where $g, \beta, \Delta \theta$ denote the gravitational acceleration, thermal expansion coefficient and temperature difference. Additional simulations at larger \(\Gamma\) (up to 1) were performed for morphological comparison in the plume condensation analysis (Fig. ~\ref{fig2}a). For statistical convergence, the simulations were integrated for a minimum of 3000 free-fall time units ($t_f$) following the initial transients. The grid resolution $\Delta_g$ is validated against the Kolmogorov scale $\eta_k$, which represents the smallest scale requiring resolution at Pr $< 1$ \citep{silanoNumericalSimulationsRayleigh2010, chongEffectPrandtlNumber2018}. Additionally, sufficient resolution within the boundary layers is ensured, following the criteria established in previous studies \citep{shishkinaBoundaryLayerStructure2010a}.

\begin{figure}
	\centering
	\includegraphics[width=1.0\linewidth]{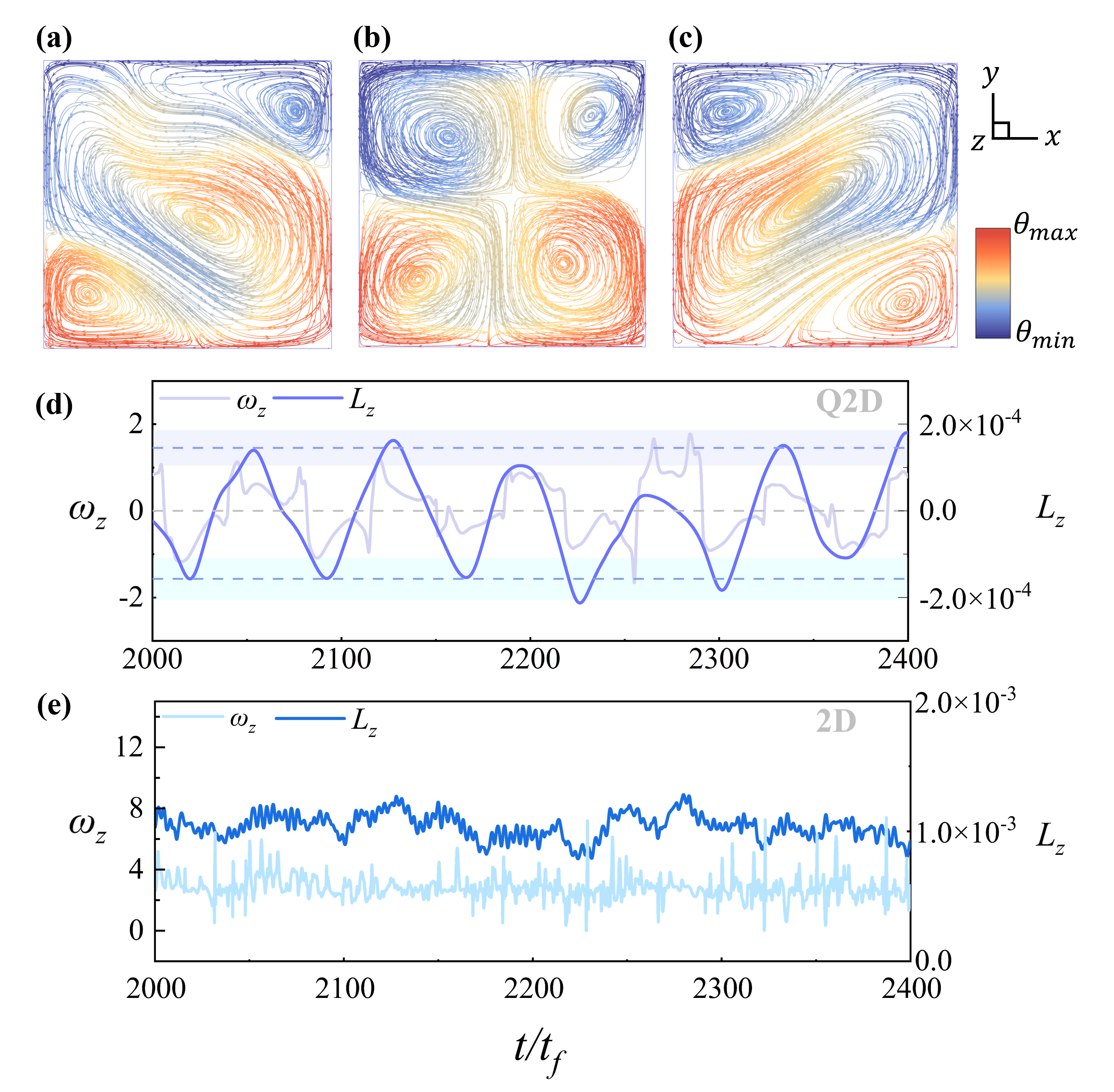}
	\caption{Dynamics of the LSC reversal at Ra$=9 \times 10^6$ and $\Gamma=0.04$. (a)-(c) Instantaneous streamlines colored by temperature showing the reversal sequence: (a) initial counter-clockwise LSC; (b) transient quadrupolar state; (c) final clockwise LSC. Time evolution of center spanwise vorticity $\omega_{z}$ and global angular momentum $L_{z}$ for (d) the quasi-2D case, showing sustained quasi-periodic reversals, and (e) the purely 2D case, which remains stable.}
    \label{fig1}
\end{figure}

\begin{figure*}[t]
	\centering
	\includegraphics[width=\textwidth]{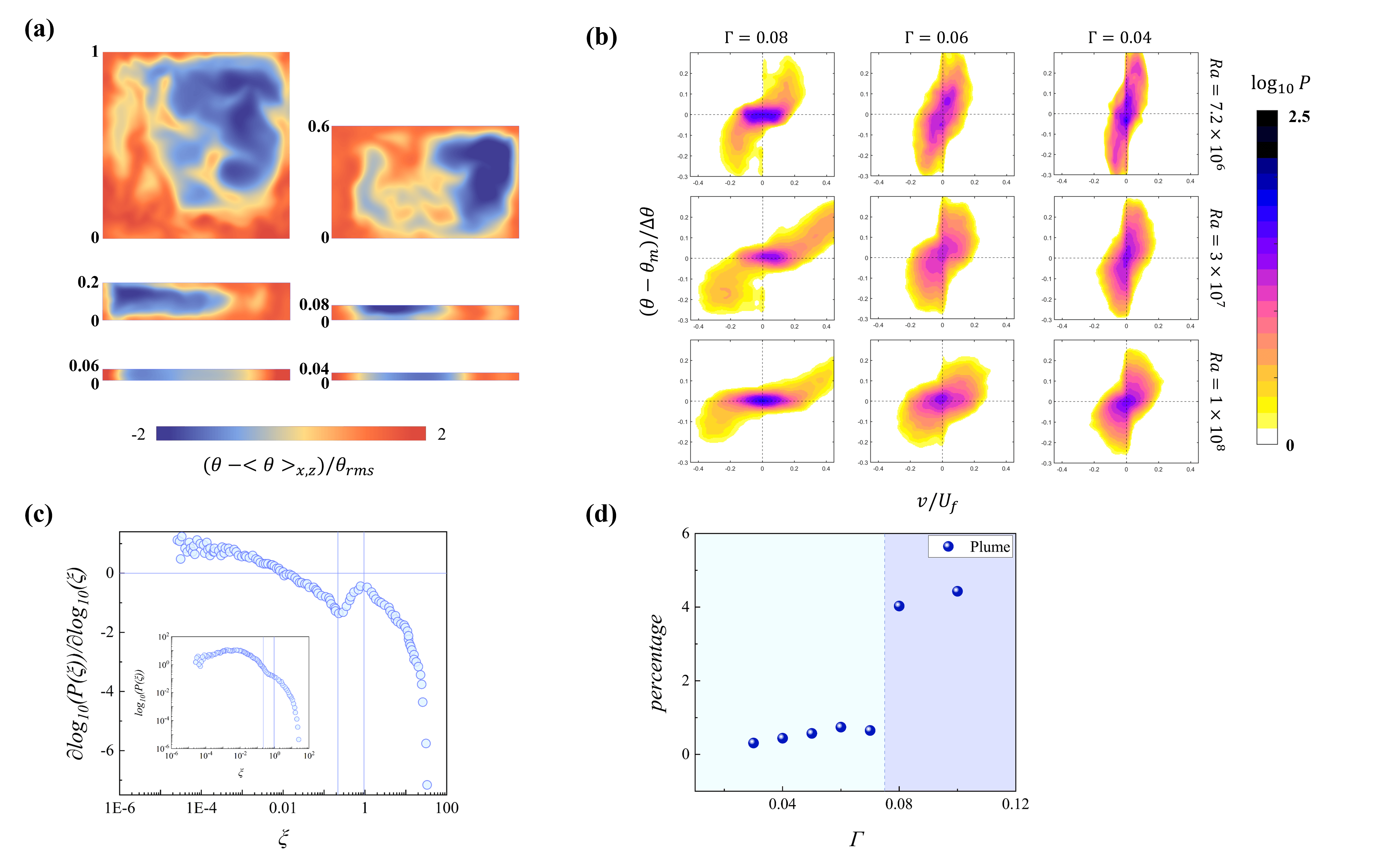}
	\caption{Confinement-induced ``plume condensation" at Ra$=9\times10^{6}$. (a) Horizontal cuts of instantaneous standardized temperature field near the boundary layer for decreasing aspect ratios $\Gamma \in [0.04,1]$, revealing a transition from chaotic clusters to coherent quasi-linear structures. (b) JPDFs of vertical velocity $v$ and temperature $\theta$ on the horizontal mid-plane ($y/H=0.5$). Strong confinement ($\Gamma=0.04$) induces compact velocity distributions and extreme temperature distributions. (c) Partitioning of thermal dissipation via the local slope of $P(\xi)$ ($\xi=\epsilon_{\theta}/\langle\epsilon_{\theta}\rangle_{V}$). (d) Volume-averaged plume dissipation fraction $p_{plume}$ vs $\Gamma$.}
    \label{fig2}
\end{figure*}

The reversal dynamics unfold through a recurring cycle that typically begins with the system dominated by a single diagonal LSC (Fig.~\ref{fig1}a). As the flow evolves, corner vortices gradually accumulate energy by capturing plumes that detach from the boundary layers, which leads to a rapid growth in both their size and kinetic energy. These expanding vortices progressively disrupt the main circulation, driving the system into a transient quadrupolar topological state (Fig.~\ref{fig1}b). This destabilization ultimately causes the original LSC to collapse, establishing a new, oppositely directed circulation (Fig.~\ref{fig1}c; see Supplementary Material for the full dynamics). Remarkably, this physical pathway—triggered via corner vortex growth and destabilization—aligns qualitatively with the mechanism previously reported for moderate Pr systems \citep{sugiyamaFlowReversalsThermally2010a}.

We quantitatively characterize the LSC reversal dynamics by monitoring the global angular momentum:
\begin{equation}
    L_z(t)=\int_{\mathcal{D}}{(\mathbf{r} \times \mathbf{u}(t,\mathbf{r}))\cdot \mathbf{e}_3\mathrm{d}A} = -\frac{1}{2}\int_{\mathcal{D}}{r^2\omega_z (t,\mathbf{r})\:\mathrm{d}A},
    \label{Lz}
\end{equation}
and the spanwise vorticity $\omega_z$ at the cell center $(L/2, H/2, W/2)$. The sign of these metrics tracks the macroscopic LSC direction \citep{podvinLargescaleInvestigationWind2015, castillo-castellanosCessationReversalsLargescale2019}.  As shown in Fig.~\ref{fig1}(d), in the strongly confined quasi-2D system, both $L_z(t)$ and $\omega_z$ exhibit pronounced and quasi-periodic oscillations, alternating between positive and negative values. This confirms systematic, global LSC reversals. The oscillating signal is approximately symmetric about zero, with the mean and variance of $L_z$ being nearly identical for the clockwise and counter-clockwise LSC states. To elucidate the critical role of sidewall confinement, we performed a purely 2D simulation at the same Ra and $L/H=1$ for comparison (Fig.~\ref{fig1}e). In the 2D system, $L_z$ and $\omega_z$ exhibit turbulent fluctuations around non-zero mean, never crossing the zero line. The LSC thus remains stable and does not reverse. This stark contrast reveals that the 3D effects introduced by lateral walls—neglected in 2D models—are the key physical mechanism triggering LSC reversals in low Pr fluids, fundamentally altering the large-scale flow dynamics.

To elucidate the mechanism underlying confinement-induced reversals, we examine how the aspect ratio $\Gamma$ modulates the internal flow structure. We find that sidewall confinement fundamentally reshapes the thermal plume morphology. As illustrated in Fig.~\ref{fig2}(a)—where horizontal cross-sections of the instantaneous standardized temperature field are visualized near the thermal boundary layer following the approach of Chong et al. \citep{chongCondensationCoherentStructures2015}—a distinct morphological transition is observed. Specifically, as $\Gamma$ is reduced below 0.08, the plumes reorganize from disordered, chaotic clusters into highly coherent, quasi-linear structures. Crucially, this ``plume condensation" is not merely a local boundary layer effect but a global flow reorganization coincident with the onset of reversals. This structural coherence dictates the statistical behavior of the flow, as evidenced by the Joint Probability Density Functions (JPDFs) of vertical velocity and temperature shown in Fig.~\ref{fig2}(b). Consistent with the findings in confined convection, the JPDFs reveal that strong confinement compacts the velocity distribution due to enhanced sidewall friction, while simultaneously elongating the temperature distribution. This elongation signifies that the condensed plumes maintain high thermal coherence against the rapid diffusion characteristic of low-Pr fluids \citep{xiaTuningHeatTransport2023}.

To quantify the impact of this coherence on energy transport, we employ a Probability Density Function (PDF) analysis of the thermal dissipation rates $\epsilon_{\theta}$, adopting the method from previous studies \citep{kaczorowskiAnalysisThermalPlumes2009}. The instantaneous $\epsilon_{\theta}$ field is normalized by its volume average, $\xi=\epsilon_{\theta}/\langle\epsilon_{\theta}\rangle_{V}$, and $P(\xi)$ is computed using exponentially-sized bins. By analyzing the local slope $|\partial \log_{10}(P(\xi))/\partial \log_{10}(\xi)|$ in a log-log plot to identify inflection points (Fig.~\ref{fig2}c), the flow field is statistically partitioned into three distinct regions: plumes, boundary layers, and the bulk. The thermal dissipation analysis (Fig.~\ref{fig2}d) reveals a decisive energetic shift: as $\Gamma$ approaches the critical value for reversals, the fractional dissipation within the plumes ($p_{plume}$) drops precipitously. This implies that the coherent, quasi-linear plumes experience significantly reduced thermal dissipation while traversing the bulk.  Such enhanced preservation of thermal energy enables the plumes to retain sufficient energy to drive the corner vortices, thereby fueling vortex growth and initiating a reversal, effectively overcoming the high-diffusivity barrier that typically suppresses reversals in low-Pr fluids.

\begin{figure}
	\centering
	\includegraphics[width=1.0\linewidth]{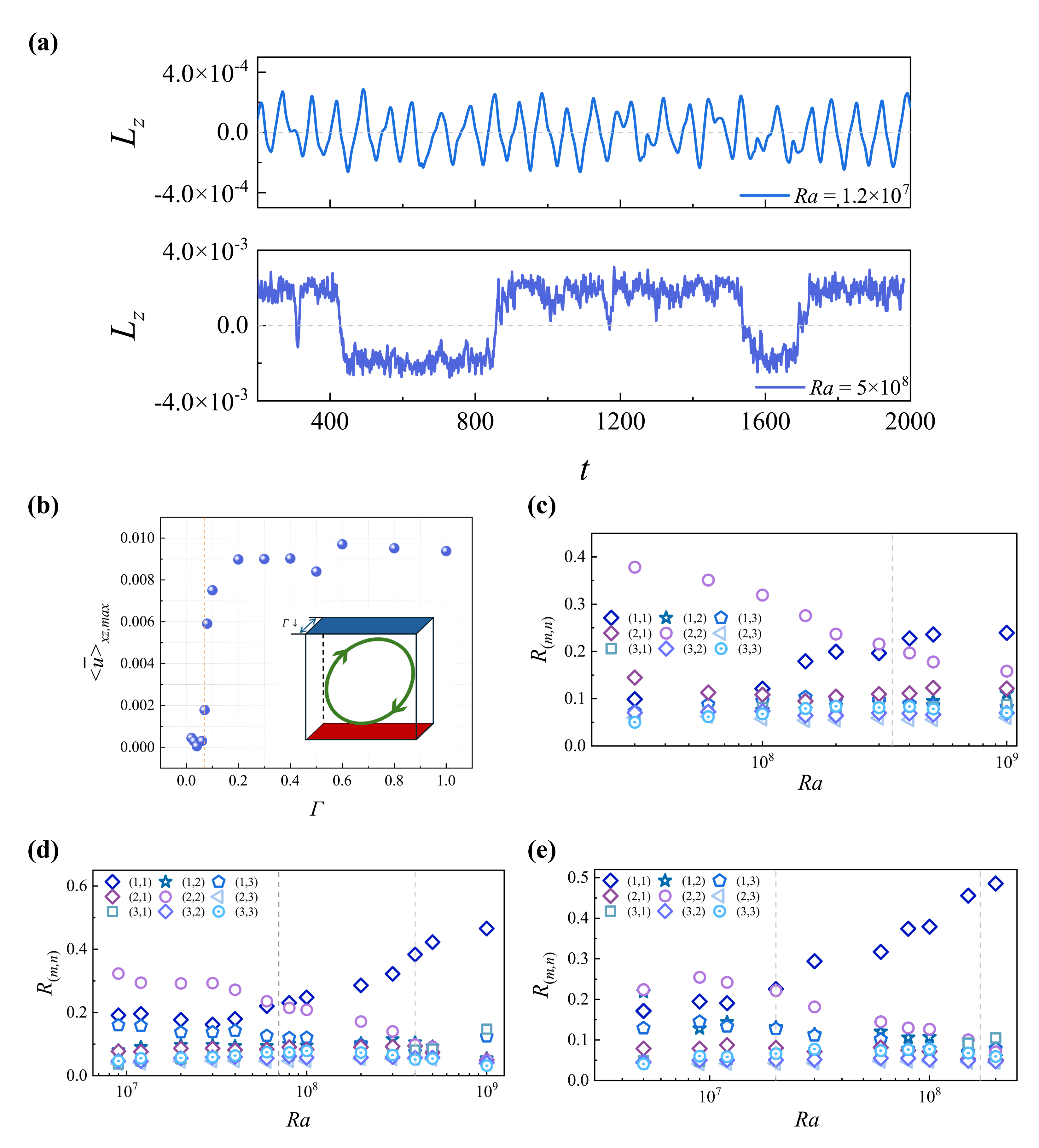}
	\caption{ Modulation of LSC dynamics by confinement $\Gamma$ and Ra. (a) Time series of $L_{z}(t)$ at $\Gamma=0.04$, showing a shift from quasi-periodic (Ra$=1.2\times10^{7}$) to intermittent reversals (Ra$=5\times10^{8}$). (b) Characteristic LSC velocity $\langle\bar{u}\rangle_{xy,max}$ vs $\Gamma$ at Ra$=9\times10^{6}$; strong confinement significantly decelerates the flow. Inset: Quasi-2D LSC schematic. (c)-(e) Energy fractions $R_{(m,n)}$ of dominant Fourier modes vs Ra for $\Gamma=0.06, 0.04, 0.02$. The crossover where the (1,1) mode surpasses the (2,2) mode correlates with the transition to intermittent behavior.}
    \label{fig3}
\end{figure}

In addition to reshaping plumes, sidewall confinement plays a further critical role by directly modulating the dynamics of the LSC. The temporal evolution of the flow, as evidenced by the time series of global angular momentum in Fig.~\ref{fig3}(a), reveals a clear transition in reversal morphology: the system shifts from quasi-periodic oscillations at lower Ra to intermittent reversals as Ra increases. This sensitivity to parameters is rooted in how confinement alters the flow's stability. As shown in Fig.~\ref{fig3}(b), the characteristic LSC velocity, $\langle \overline{u} \rangle_{xy, max}$, exhibits a marked deceleration as the aspect ratio $\Gamma$ is reduced. While the damping of reversals is typically attributed to the stabilization of corner vortices \citep{niReversalsLargescaleCirculation2015a}, this confinement-induced deceleration fundamentally alters the stability landscape. By reducing the inertial momentum of the main circulation, confinement renders the LSC less robust and more susceptible to perturbations from growing corner vortices, thereby predisposing the system to reversals. In this strongly confined regime, heat transport is no longer restricted to the periphery but becomes active throughout the central region \citep{huangConfinementInducedHeatTransportEnhancement2013}. Consequently, the flow topology deviates from the regular structure observed at lower Ra (e.g., Fig.~\ref{fig1}a) and becomes increasingly populated by small-scale vortex structures. These vortices originate from energetic plumes that detach vigorously from the thermal boundary layers and subsequently stagnate within the cavity due to enhanced sidewall friction. To disentangle these complex turbulent dynamics, we employ Fourier mode decomposition, a robust method for analyzing large-scale flow structures in turbulent RB convection \citep{xuCorrelationInternalFlow2020, chandraFlowReversalsTurbulent2013, gaoFlowStateTransition2024}. Specifically, the instantaneous velocity fields are projected onto the Fourier basis $u(x, y, t) = \sum_{m,n} \hat{u}(m, n, t) [2 \sin(m \pi x) \cos(n \pi y)]$ and $v(x, y, t) = \sum_{m,n} \hat{v}(m, n, t) [-2 \cos(m \pi x) \sin(n \pi y)]$. The modal energy $E^{m,n}(t)=\sqrt{[A_{x}^{m,n}(t)]^{2}+[A_{y}^{m,n}(t)]^{2}}$ (where $A_{x}^{m,n}(t)=\langle u(x,y,t),\hat{u}^{m,n}(x,y)\rangle$ and $A_{y}^{m,n}(t)=\langle v(x,y,t),\hat{v}^{m,n}(x,y)\rangle$) reveals that the entire reversal process is dominated by a competition between the (1,1) LSC mode and the (2,2) quadrupolar transition state. Figures~\ref{fig3}(c-e) show the evolution of the modal energy fraction, $R_{(m,n)} = \langle E^{m,n}(t)\rangle/\langle \sum_{m,n}E^{m,n}(t)\rangle $, versus Ra for different $\Gamma$. As Ra increases, a clear crossover in mode dominance is observed: the energy fraction of the LSC (1,1) mode, $R_{(1,1)}$, steadily rises, whereas that of the (2,2) transition mode, $R_{(2,2)}$, declines. This shifting energy balance directly dictates the reversal morphology. In the low Ra regime, $R_{(2,2)}$ is notably higher than $R_{(1,1)}$, indicating that the quadrupolar state is energetically favored; this predisposes the system to frequent, quasi-periodic reversals (Fig.~\ref{fig3}d). Conversely, at higher Ra, $R_{(1,1)}$ surpasses $R_{(2,2)}$ and becomes the dominant mode, resulting in a more stable LSC that is only intermittently disrupted (Fig.~\ref{fig3}e). Eventually, at sufficiently high Ra, $R_{(2,2)}$ mode becomes negligible as other modes gain prominence, thereby preventing the formation of a transition state and suppressing reversals.

\begin{figure}
	\centering
	\includegraphics[width=1.0\linewidth]{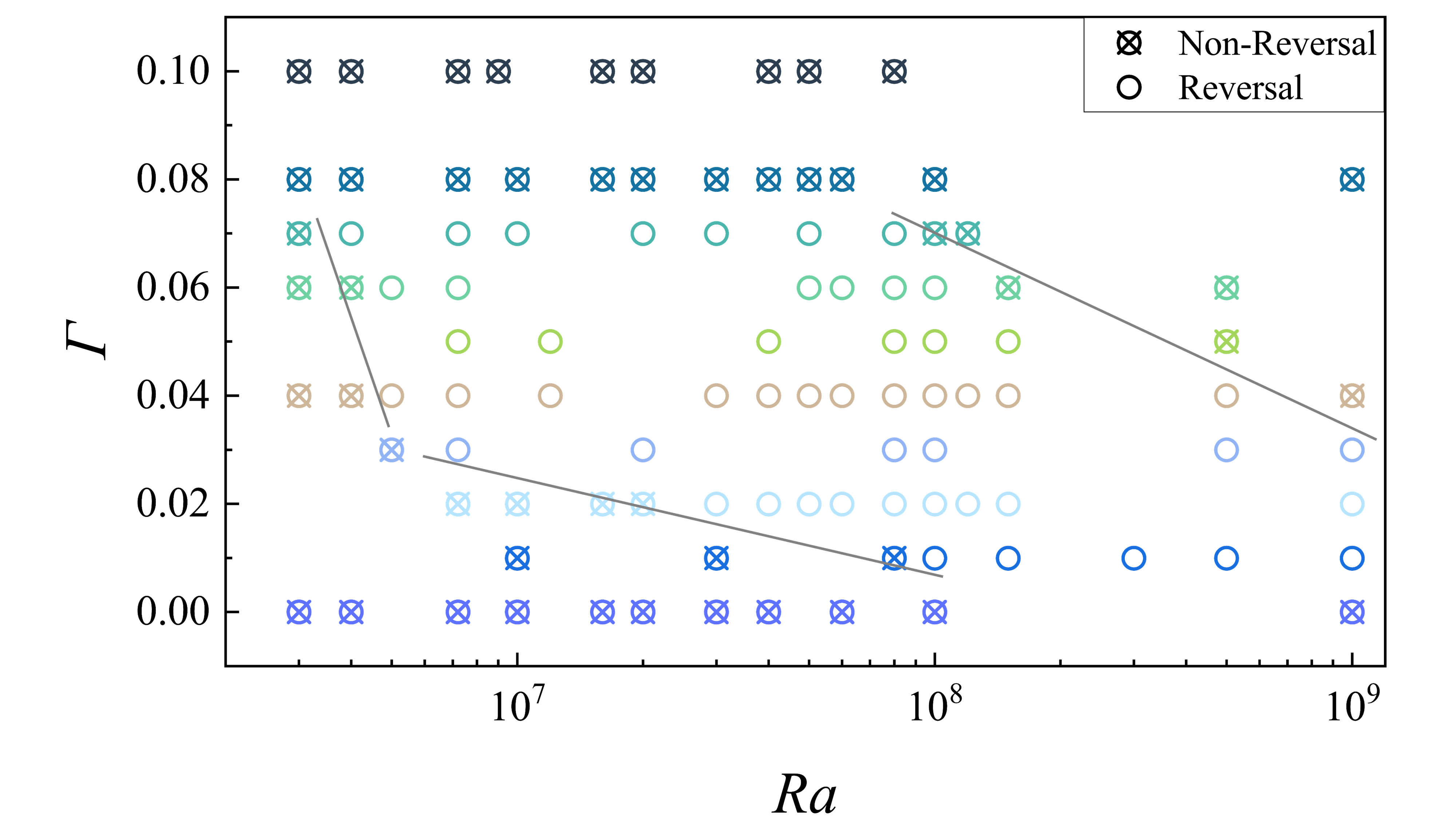}
	\caption{Phase diagram in the Ra-$\Gamma$ space for Pr$=0.029$. Open circles ($\circ$): sustained LSC reversals; crossed circles ($\otimes$): non-reversing flows. Solid lines delineate the ``island of reversal", a regime where strong lateral confinement-induced plume condensation overcomes high thermal diffusivity to enable reversals.}
    \label{fig4}
\end{figure}

The $\Gamma-$Ra phase diagram (Fig.~\ref{fig4}) synthesizes these findings, delineating a specific ``island of reversal" where LSC reversals are sustained. This reversal zone is bounded by three distinct physical mechanisms: (i) At the low Ra boundary, viscous friction dominates under strong confinement, rendering the buoyant driving force insufficient to trigger the necessary corner vortex instability. (ii) At the high Ra boundary, the LSC becomes increasingly robust and inertially stable, a trend consistent with the stretched exponential dependence of reversal time on Ra reported by \citet{niReversalsLargescaleCirculation2015a}; this enhanced stability suppresses corner vortex intrusion and thereby prevents reversals. (iii) The high-$\Gamma$ boundary, which is Ra-dependent (see Fig.~\ref{fig4}), marks the limit where confinement effects weaken. Beyond this boundary, the system transitions to a state of weaker confinement, where the flow exhibits increased three-dimensionality with chaotic plumes and elevated thermal dissipation. This inhibits the effective energy accumulation by corner vortices necessary to trigger a reversal. Consequently, the realization of reversals in low Pr fluids relies on a delicate balance: sidewall confinement must be sufficiently strong (small $\Gamma$) to organize coherent, low-thermal-dissipation plumes, yet not so excessive that viscous damping suppresses the underlying dynamical instabilities.

In summary, we have uncovered an ``island of reversal" for low-$Pr$ fluids, a regime previously thought to be dynamically stable. The mechanism hinges on lateral confinement, which induces a plume condensation transition that overcomes the barrier of high thermal diffusivity while simultaneously reducing inertial stability via sidewall friction. This confinement-controlled regime demonstrates that geometric constraints can be as potent as fluid properties in dictating turbulent states. Our findings suggest that strong lateral confinement serves as an effective control parameter to manipulate coherent structures, potentially offering a general mechanism to trigger flow instabilities in other turbulent systems where rapid thermal diffusion typically suppresses convective dynamics.

\begin{acknowledgments}
The authors acknowledge support from NSFC under grant No.52522605, U24A6007, CAS under grant No.YSBR-087, XDB0790103, and the National Key R\&D Programme of China (2022YFA1204100).
\end{acknowledgments}

\bibliographystyle{apsrev4-2} 
\bibliography{Reversal}

\end{document}